\documentclass[pra,twocolumn,superscriptaddress,10pt]{revtex4-1}
\usepackage{graphicx}
\usepackage{dcolumn}
\usepackage{color}
\usepackage{times}
\usepackage{bm}
\usepackage{amssymb}
\usepackage{amsmath}
\usepackage{epsfig}
\usepackage{epstopdf}
\usepackage{dsfont}
\usepackage{subfigure}
\begin{document}
	
	\title{\textbf{Experimental investigation of nonlocal advantage of quantum coherence}}
	
	\author{Zhi-Yong Ding}
	\affiliation{School of Physics and Material Science, Anhui University, Hefei 230601, China}
	\affiliation{School of Physics and Electronic Engineering, Fuyang Normal University, Fuyang 236037, China}
	\affiliation{Key Laboratory of Functional Materials and Devices for Informatics of Anhui Educational Institutions, Fuyang Normal University, Fuyang 236037, China}

	\author{Huan Yang}
	\affiliation{School of Physics and Material Science, Anhui University, Hefei 230601, China}
	\affiliation{Department of Experiment and Practical Training Management, West Anhui University, Lu'an 237012, China}
		
	\author{Hao Yuan}
	\email[]{hyuan@ustc.edu.cn (H. Yuan)}
	\affiliation{School of Physics and Material Science, Anhui University, Hefei 230601, China}
	\affiliation{CAS Key Laboratory of Quantum Information, University of Science and Technology of China, Hefei 230026, China}
	
	\author{Dong Wang}
	\email[]{dwang@ahu.edu.cn (D. Wang)}
	\affiliation{School of Physics and Material Science, Anhui University, Hefei 230601, China}
	\affiliation{CAS Key Laboratory of Quantum Information, University of Science and Technology of China, Hefei 230026, China}
	
	\author{Jie Yang}
	\affiliation{School of Physics and Material Science, Anhui University, Hefei 230601, China}
	
	\author{Liu Ye}
	\email[]{yeliu@ahu.edu.cn (L. Ye)}
	\affiliation{School of Physics and Material Science, Anhui University, Hefei 230601, China}
	
	\begin{abstract}
		Nonlocal advantage of quantum coherence (NAQC) based on coherence complementarity relations is generally viewed as a stronger nonclassical correlation than Bell nonlocality. An arbitrary two-qubit state with NAQC must be an entangled state, which demonstrates that the criterion of NAQC can also be regarded as an entanglement witness. In this paper, we experimentally investigate the NAQC for Bell-diagonal states with high fidelity in an optics-based platform. We perform local measurements on a subsystem in three mutually unbiased bases and reconstruct the density matrices of the measured states by quantum state tomography process. By analyzing characteristic of the $l_1$ norm, relative entropy and skew information of coherence with parameters of quantum states, NAQC for the quantum states is accurately captured, and it shows that our experimental results are well compatible with the theoretical predictions. It is worth mentioning that quantum states with NAQC would have higher entanglement, and thus NAQC could be expected to be a kind of useful physical resource for quantum information processing.
	\end{abstract}
	\date{\today}
	\maketitle
	
	\section{INTRODUCTION}
		Nonclassical correlations, being the most incredible feature of the quantum world, can enable classically impossible tasks \cite{w01}. So far, for two-qubit states, a distinct hierarchy of these typical correlations, including Bell nonlocality \cite{w02}, quantum entanglement \cite{w03}, quantum steering \cite{w04}, and quantum discord \cite{w05} has been gradually clarified \cite{w06, w07, w08, w09, w10}. From the application perspective, these correlations can also be regarded as valuable physical resource applied to quantum information processing tasks \cite{w11, w12}, such as device-independent quantum cryptography \cite{w13, w14, w15, w16, w17}, quantum teleportation and dense coding \cite{w18, w19, w20, w21}, quantum metrology \cite{w22, w23, w24} and so on. These abundant applications not only deepen our understanding of quantum information theory, but also become an important motivation for making efforts to facilitate the advance of this field. On the other hand, coherence \cite{w25, w26}, as one of the most fundamental features of quantum system, originates from the superposition principle in quantum mechanics. To quantify coherence of quantum states, Baumgratz \textit{et al}. \cite{w27} formulated a rigorous groundbreaking framework for defining a valid coherence measure and proposed two distance-based coherence measures, i.e., the $l_1$ norm and relative entropy of coherence. In addition, other faithful coherence measures have been proposed, such as the entanglement-based measure of coherence \cite{w28}, the intrinsic randomness of coherence \cite{w29}, the robustness of coherence \cite{w30, w31}, and the skew information of coherence \cite{w32, w33, w34}, etc.
	
		Although coherence is usually considered as be one of the characteristics of the whole physical system, it can be associated with other nonclassical correlations based on resource theory. Even conceptually, coherence is more fundamental than other correlations \cite{w26}. For a bipartite quantum system, coherence between subsystems can be deemed as its ability to create entanglement \cite{w28}. It means that coherence can be converted to entanglement by incoherent operations applied to the system and an incoherent ancilla. Furthermore, it has been proved  that coherence and quantum discord can be transformed into each other in theoretical  and experimental aspects \cite{w35, w36}.
	
		Recently, Mondal \textit{et al}. \cite{w37} established a connection between quantum steering and coherence based on coherence complementarity relations. They showed that, for a bipartite qubit state $AB$, the average coherence of the conditional states of subsystem $B$ after performing local measurements on qubit $A$ may break through the coherence limit of a single-qubit state measured on mutually unbiased bases. Typically, Mondal \textit{et al}. called this interesting effect as ``nonlocal advantage of quantum coherence" (NAQC), and formulated the criterion for achieving NAQC by using different coherence measures. To reveal the relationship between NAQC and other quantum correlation measures, Hu \textit{et al}. \cite{w38} investigated the hierarchy of NAQC and Bell nonlocality, and testified that NAQC is a new nonclassical correlation which is stronger than Bell nonlocality. Any bipartite qubit state with NAQC must be an entangled state, which demonstrates that the criterion of NAQC can also be regarded as an entanglement witness. Further, Hu \textit{et al}. \cite{w39} generalized the framework of NAQC from two qubit states to high-dimensional states. Datta \textit{et al}. \cite{w40} provided a scenario for sharing of NAQC by sequential observers. Such studies have deepened our understanding of the quantum resource theories \cite{w11, w25}, and NAQC is expected to be a kind of useful physical resource for quantum information processing. Nevertheless, the investigations of NAQC still remain theoretical, and there is few experimental investigation of NAQC up to now.
	
		 We herein report an experimental investigation of NAQC for two-qubit states in an all-optical setup. By using quantum state tomography process to reconstruct the initial states and measured states,  we capture NAQC for the quantum states based on the $l_1$ norm, relative entropy and skew information of coherence, respectively. It can be found that NAQC depends not only on the form of quantum states, but also on the choice of coherence measures. Note that, our experimental results coincide with the theoretical predictions very well.
		
	\section{PRELIMINARIES}
		For a single qubit state described by density operator , three widely established coherence measures: the $l_1$ norm \cite{w27}, relative entropy \cite{w27} and skew information \cite{w34} of coherence in the basis of eigenvectors of the Pauli spin observable ${\sigma _i}\;(i = x,y,z)$ are given by
		\begin{subequations}
			\begin{align}
			&C_{{l_1}}^{{\sigma _i}}(\rho ) = \sum\limits_{\mu  \ne \nu } {\left| {\left\langle \mu  \right|\rho \left| \nu  \right\rangle } \right|} \\
			&C_{re}^{{\sigma _i}}(\rho ) = S({\rho ^{diag}}) - S(\rho ),\label{g01b}\\
			&C_{sk}^{{\sigma _i}}(\rho ) =  - \frac{1}{2}{\rm{Tr}}{\left[ {\sqrt \rho  ,{\sigma _i}} \right]^2},\label{g01c}
			\end{align}
		\end{subequations}
		 respectively, where $\{ \left| \mu  \right\rangle ,\left| \nu  \right\rangle \} $ are the eigenvectors of ${\sigma _i}$,  $S(\cdot)$ denotes the von Neumann entropy, and ${\rho ^{diag}}{\rm{ = }}\sum\nolimits_{\mu ,\nu } {\left| \mu  \right\rangle \left\langle \mu  \right|\left\langle \mu  \right|\rho \left| \mu  \right\rangle } $ is the completely decohered state of $\rho$. Specially, the complementarity relation of coherence under mutually unbiased bases has been derived in Ref. \cite{w37},
		\begin{equation}\label{g02}
			  \sum\limits_{i = x,y,z} {C_\alpha ^{{\sigma _i}}(\rho )}  \le C_\alpha ^m,
		\end{equation}
		where $C_\alpha ^m\;(\alpha  = {l_1},re,sk)$ are state-independent upper bound and their values are $C_{{l_1}}^m = \sqrt 6  \approx 2.45$, $C_{{re}}^m \approx 2.23$ and $C_{sk}^m = 2$. The equality is hold for a special pure state ${\rho ^{\max }} = \frac{1}{2}[\mathbb{I} + \frac{1}{{\sqrt 3 }}({\sigma _x} + {\sigma _y} + {\sigma _z})]$, where $\mathbb{I}$ is the identity operator.
		
		Based on the above coherence measures, we briefly introduce the ``steering game'' \cite{w37} proposed by Mondal \textit{et al}. and give the criterion for achieving NAQC. Assuming that Alice and Bob share a bipartite qubit state ${\rho _{AB}}$. Alice then performs a local measurement $\Pi _i^a = [\mathbb{I} + {( - 1)^a}{\sigma _i}]/2$ on her side (qubit $A$) in the eigenbasis of ${\sigma _i}$ and obtains an outcome $a = \{ 0,1\} $ with probability ${p_{\Pi _i^a}} = {\rm{Tr[}}(\Pi _i^a \otimes {\mathbb{I}_B}){\rho _{AB}}{\rm{]}}$. Then the measured states of the bipartite system can be expressed as ${\rho _{AB|\Pi _i^a}} = {(\Pi _i^a \otimes {\mathbb{I}_B}){\rho _{AB}}(\Pi _i^a \otimes {\mathbb{I}_B})} /{p_{\Pi _i^a}}$, and the ensemble of conditional state on Bob's side becomes $\{ {p_{\Pi _i^a}},{\rho _{B|\Pi _i^a}}\} $, where ${\rho _{B|\Pi _i^a}} = {\rm{T}}{{\rm{r}}_A}({\rho _{AB|\Pi _i^a}})$. Alice informs Bob of her choice of measurement and the corresponding outcome, and then Bob measures the coherence of his side (qubit $B$) randomly in the eigenbasis of either ${\sigma _j}$ or ${\sigma _k}$ ($j,k \ne i$). Since Alice can choose any of the six local measurement settings, and Bob chooses the corresponding possible reference eigenbases, the criterion for achieving NAQC can be obtained by all possible probabilistic averaging \cite{w37},
		\begin{equation}\label{g03}
			{N_\alpha }({\rho _{AB}}) = \frac{1}{2}\sum\limits_{i,j,a} {p({\rho _{\Pi _{j \ne i}^a}})C_\alpha ^{\sigma _i}({\rho _{B|\Pi _{j \ne i}^a}})}  > C_\alpha ^m.
		\end{equation}	
		
		Basically, NAQC can be seen as a stronger nonclassical correlation than Bell nonlocality \cite{w38}. To detect Bell nonlocality of a two-qubit state ${\rho _{AB}}$, we can use the well-known Clauser-Horne-Shimony-Holt (CHSH) inequalitie $\left| {\left\langle {{B_{CHSH}}} \right\rangle } \right| = \left| {{\rm{Tr}}({\rho _{AB}}{B_{CHSH}})} \right| \le 2$ \cite{w41}, where ${B_{CHSH}}$ denotes the Bell operator. Violation of CHSH inequality means that the quantum state is Bell nonlocalized. The maximum violation of CHSH inequality can be represented as \cite{w42}
		\begin{equation}\label{g04}
			{B_{\max }}({\rho _{AB}}) = 2\sqrt {m({\rho _{AB}})} ,
		\end{equation}
		where $m({\rho _{AB}}) = {\max _{i < j}}({u_i} + {u_j})$, ${u_i}\;(i = 1,2,3)$ is the eigenvalue of ${T^\dag }T$ (${T^\dag }$ stands for transposition of $T$), and $T$ is the correlation tensor in the Bloch representation with elements ${t_{ij}} = {\rm{Tr}}({\rho _{AB}}{\sigma _i} \otimes {\sigma _j})$.
		
		Furthermore, any bipartite quantum state with NAQC is an entangled state, thus the criterion of NAQC can also be regarded as an entanglement witness \cite{w39}. It is known that entanglement measure can be quantified conveniently by concurrence \cite{w43}. For a two-qubit state ${\rho _{AB}}$, the concurrence is defined as	
		\begin{equation}\label{g05}
			C({\rho _{AB}}) = \max \{ 0,{\lambda _1} - {\lambda _2} - {\lambda _3} - {\lambda _4}\},
		\end{equation}
		where ${\lambda _1}, \cdots ,{\lambda _4}$ are the square roots of the eigenvalues of the non-Hermitian matrix ${\rho _{AB}}({\sigma _y} \otimes {\sigma _y})\rho _{AB}^*({\sigma _y} \otimes {\sigma _y})$. The variable $\rho _{AB}^*$ is the complex conjugate of ${\rho _{AB}}$ in the fixed basis $\{ \left| {00} \right\rangle ,\left| {01} \right\rangle ,\left| {10} \right\rangle ,\left| {11} \right\rangle \}$.
		
	\section{EXPERIMENTS}
		To experimentally investigate NAQC shown in Eq. (3), we first need to prepare a bipartite qubit state with high fidelity and then perform six different measurement settings $\{ \Pi _i^a\} $ on qubit $A$ in the eigenbasis of ${\sigma _i}\;(i = x,y,z)$. In this paper, we choose a Bell-diagonal state (BDS) \cite{w44, w45} $\rho _{AB}$ as the initial state with form of,
		\begin{equation}\label{g06}
			\begin{split}
				{\rho _{AB}} &= a\left| {{\Phi ^ + }} \right\rangle \left\langle {{\Phi ^ + }} \right| + b\left| {{\Phi ^ - }} \right\rangle \left\langle {{\Phi ^ - }} \right|\\
				&+ c\left| {{\Psi ^ + }} \right\rangle \left\langle {{\Psi ^ + }} \right| + d\left| {{\Psi ^ - }} \right\rangle \left\langle {{\Psi ^ - }} \right|,
			\end{split}
		\end{equation}
		where $\left| {{\Phi ^ \pm }} \right\rangle  = 1/\sqrt 2 (\left| {00} \right\rangle  \pm \left| {11} \right\rangle )$  and  $\left| {{\Psi ^ \pm }} \right\rangle  = 1/\sqrt 2 (\left| {01} \right\rangle  \pm \left| {10} \right\rangle )$ denote four Bell states and the  coefficients satisfy the normalization condition $a + b + c + d = 1$.
		
		\begin{figure}[t]
			\centering
			\includegraphics[width=8cm]{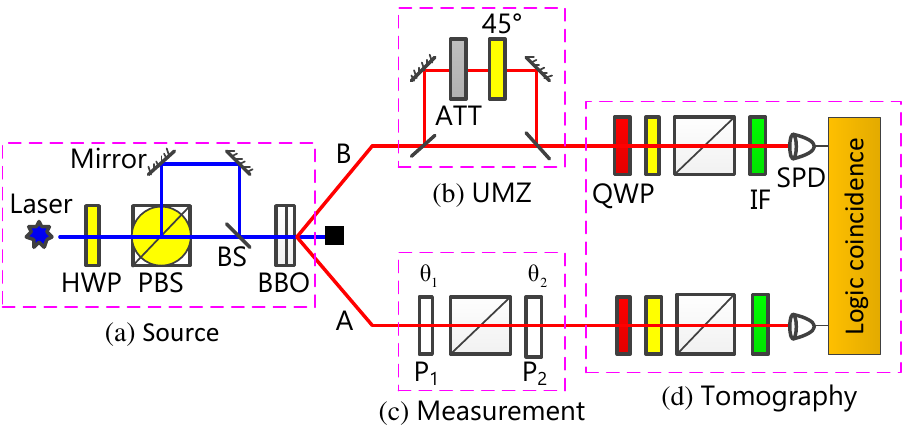}\\
			\caption{Experimental setup. (a) Source module, which is used to generate mixed polarization-entangled photon pairs, mainly includes a high-power tunable diode laser, three half-wave plates (HWPs), a polarization beam splitter (PBS), a beam splitter (BS), two mirrors, and two type-I $\beta$-barium borate (BBO) crystals. (b) UMZ module, whose purpose is to achieve the proportionate mixing of quantum states, includes an attenuator (ATT) , a HWP, two mirrors, and two BSs. (c) Measurement module, which is use to realize local measurement on the photon A, consists of two wave plates (${P}_1$ and $P_2$) and a PBS. (d) Tomography module, whose function is to reconstruct quantum states, includes two quarter-wave plates (QWPs), two HWPs, two PBSs, two 3-nm interference filter (IFs) , two single-photon detector (SPDs), and a dual-channel logic coincidence meter.}\label{Fig1}
		\end{figure}
		
		Technically, the polarizations of photons can be encoded as information carriers in the experimental investigation of quantum information processing tasks \cite{w12}. In the current experiment, we encode the horizontal polarization state $\left| H \right\rangle $ as $\left| 0 \right\rangle $ and the vertical polarization state $\left| V \right\rangle $ as  $\left| 1 \right\rangle $. Explicitly, the optical experimental setup has been illustrated in  Fig. 1. The design consists of  four modules represented by the purple dotted frames, respectively: (a) Source module, (b) Unbalanced Mach-Zehnder device (UMZ) module, (c) Measurement module and (d) Tomography module. The function of module (a) is to generate mixed polarization-based entangled photon pairs. A high-power linearly polarized pumped beam (130mW, 405nm) emitted by the tunable diode laser is divided into two parts by the polarization beam splitter (PBS). The transmitted beam is  $45 ^{\circ}$ linearly polarized $1/\sqrt 2 (\left| H \right\rangle  + \left| V \right\rangle )$, and the reflected beam is $-45 ^{\circ}$ linearly polarized $1/\sqrt 2 (\left| H \right\rangle  - \left| V \right\rangle )$. The relative power between these two parts can be easily adjusted by changing the angle of optical axis of the half-wave plate (HWP), and the corresponding proportional coefficient is expressed by parameter $p$. Time difference between them can be eliminated by adjusting the time-delay. The transmitted beam and reflected beam recombine by a beam splitter (BS). Then the mixed polarization-entangled photon pairs $p\left| {{\Phi ^ + }} \right\rangle \left\langle {{\Phi ^ + }} \right| + (1 - p)\left| {{\Phi ^ - }} \right\rangle \left\langle {{\Phi ^ - }} \right|$ are generated by the spontaneous parametric down-conversion (SPDC) process of the pumped beam passing through two type-I $\beta$-barium borate (BBO) crystals ($6.0 \times 6.0 \times 0.5$mm) with optic axis cut at $29.2 ^{\circ}$ \cite{w46}. In order to achieve the mixing of four Bell states as Eq. (6), we pass parametric photons in the $B$-path through the UMZ module \cite{w47}. The attenuator (ATT) is used to control the proportionality coefficient in the state. Finally, the initial BDS can be prepared as
		\begin{equation}\label{g07}
			\begin{split}
				{\rho _{AB}} &= p\left[ {q\left( {\left| {{\Phi ^ + }} \right\rangle \left\langle {{\Phi ^ + }} \right|} \right) + \left( {1 - q} \right)\left( {\left| {{\Psi ^ + }} \right\rangle \left\langle {{\Psi ^ + }} \right|} \right)} \right]\\
				&+ \left( {1 - p} \right)\left[ {q\left( {\left| {{\Phi ^ - }} \right\rangle \left\langle {{\Phi ^ - }} \right|} \right) + \left( {1 - q} \right)\left( {\left| {{\Psi ^ - }} \right\rangle \left\langle {{\Psi ^ - }} \right|} \right)} \right].
			\end{split}
		\end{equation}		
		In module (c), two wave plates ($P_1$ and $P_2$) and a PBS, and it is use to realize local measurement on the photon $A$. Six different measurement settings can be achieved by choosing the type of wave plates and rotating the angles ($\theta_1$ and $\theta_2$) of optical axis of wave plates (see Table I for details). The function of module (d) is to realize the tomography of quantum states. Each path of this module includes a quarter-wave plate (QWP), a HWP, a PBS, a 3-nm interference filter (IF) and a single-photon detector (SPD). The photons detected by the SPDs are sent to the coincidence meter for logic coincidence calculation. By changing angles of the QWPs and HWPs and recording the measured coincidence counts under at least 16 typical measurement bases \cite{w48}, the density matrix of the prepared state can be reconstructed. With the tomography module, both the initial BDS ${\rho _{AB}}$ and the measured states $\{ {\rho _{AB|\Pi _i^a}}\} $ can be reconstructed faithfully. In particular, the probability of the measured states ${p_{\Pi _i^a}}$ is also calculated by the measured coincidence counts \cite{w48}.
		
		\begin{table}[b]
			\centering
			\caption{Types and angles of wave plates for different local measurement settings on the photon $A$.}
			\begin{ruledtabular}
				\begin{tabular}{ccccc}
					Measurement & $P_1$ & $\theta_1$ & $P_2$ & $\theta_2$\\
					\colrule
					$\Pi _x^0$ & HWP & $22.5 ^{\circ}$ & HWP & $22.5 ^{\circ}$\\
					$\Pi _x^1$ & HWP & $-22.5 ^{\circ}$ & HWP & $-22.5 ^{\circ}$\\
					$\Pi _y^0$ & QWP & $45 ^{\circ}$ & QWP & $-45 ^{\circ}$\\
					$\Pi _y^1$ & QWP & $-45 ^{\circ}$ & QWP & $45 ^{\circ}$\\
					$\Pi _z^0$ & HWP & $0 ^{\circ}$ & HWP & $0 ^{\circ}$\\
					$\Pi _z^1$ & HWP & $45 ^{\circ}$ & HWP & $45 ^{\circ}$\\
				\end{tabular}
			\end{ruledtabular}		
		\end{table}

		In the following experiments, we choose different state parameters $p$ ($p$ = 0, 0.05, 0.95, 1) and keep them unchanged. By adjusting another state parameter $q$ ($q$ = 0, 0.05, 0.3, 0.5, 0.7, 0.95, 1), we can prepare various initial BDSs with different entanglement. Usually, the preparation quality of a quantum state $\rho$ can be measured by the fidelity of it and the goal state ${\rho _0}$, i.e.,  $F(\rho ,{\rho _0}) \equiv {\rm{Tr}}\sqrt {\sqrt \rho  {\rho _0}\sqrt \rho  } $ \cite{w12}. To ensure the high fidelity of the prepared initial BDSs, according to the designed experimental scheme, we need to adjust the parameters $p$ and $q$ for preparing four groups of 28 initial states (see Table II for specific parameter settings). We reconstruct their density matrices by quantum state tomography process \cite{w48} and calculate the fidelity, respectively. The average fidelity is $\bar F = 0.9978 \pm 0.0010$, where the errorbar is estimated according to the Poisson distribution characteristics of the laser. Table II shows the fidelity of the prepared states and the goal states with the different state parameters $p$ and $q$.

		\begin{table*}
			\small
			\centering
			\caption{Fidelity of the prepared states and the goal states with different state parameters $p$ and $q$.}
			\begin{ruledtabular}
				\begin{tabular}{lccccccc}
					Fidelity & $q$ = 0 & $q$ = 0.05 & $q$ = 0.3 & $q$ = 0.5 & $q$ = 0.7 & $q$ = 0.95 & $q$ = 1\\
					\colrule
					$p$ = 0 & 0.9979 $\pm$ 0.0006 & 0.9967 $\pm$ 0.0013 & 0.9969 $\pm$ 0.0013 & 0.9970 $\pm$ 0.0014 & 0.9970 $\pm$ 0.0011 & 0.9968 $\pm$ 0.0014 & 0.9975 $\pm$ 0.0003 \\
					$p$ = 0.05 & 0.9993 $\pm$ 0.0001 & 0.9952 $\pm$ 0.0012 & 0.9981 $\pm$ 0.0007 & 0.9984 $\pm$ 0.0008 & 0.9981 $\pm$ 0.0015 & 0.9986 $\pm$ 0.0003 & 0.9995 $\pm$ 0.0001 \\
					$p$ = 0.95 & 0.9992 $\pm$ 0.0001 & 0.9992 $\pm$ 0.0008 & 0.9998 $\pm$ 0.0002 & 0.9998 $\pm$ 0.0003 & 0.9997 $\pm$ 0.0003 & 0.9990 $\pm$ 0.0008 & 0.9994 $\pm$ 0.0001 \\
					$p$ = 1 & 0.9975 $\pm$ 0.0007 & 0.9944 $\pm$ 0.0024 & 0.9953 $\pm$ 0.0019 & 0.9960 $\pm$ 0.0017 & 0.9967 $\pm$ 0.0018 & 0.9971 $\pm$ 0.0021 & 0.9986 $\pm$ 0.0012 \\
				\end{tabular}
			\end{ruledtabular}		
		\end{table*}

		\begin{figure}
			\centering
			\includegraphics[width=8cm]{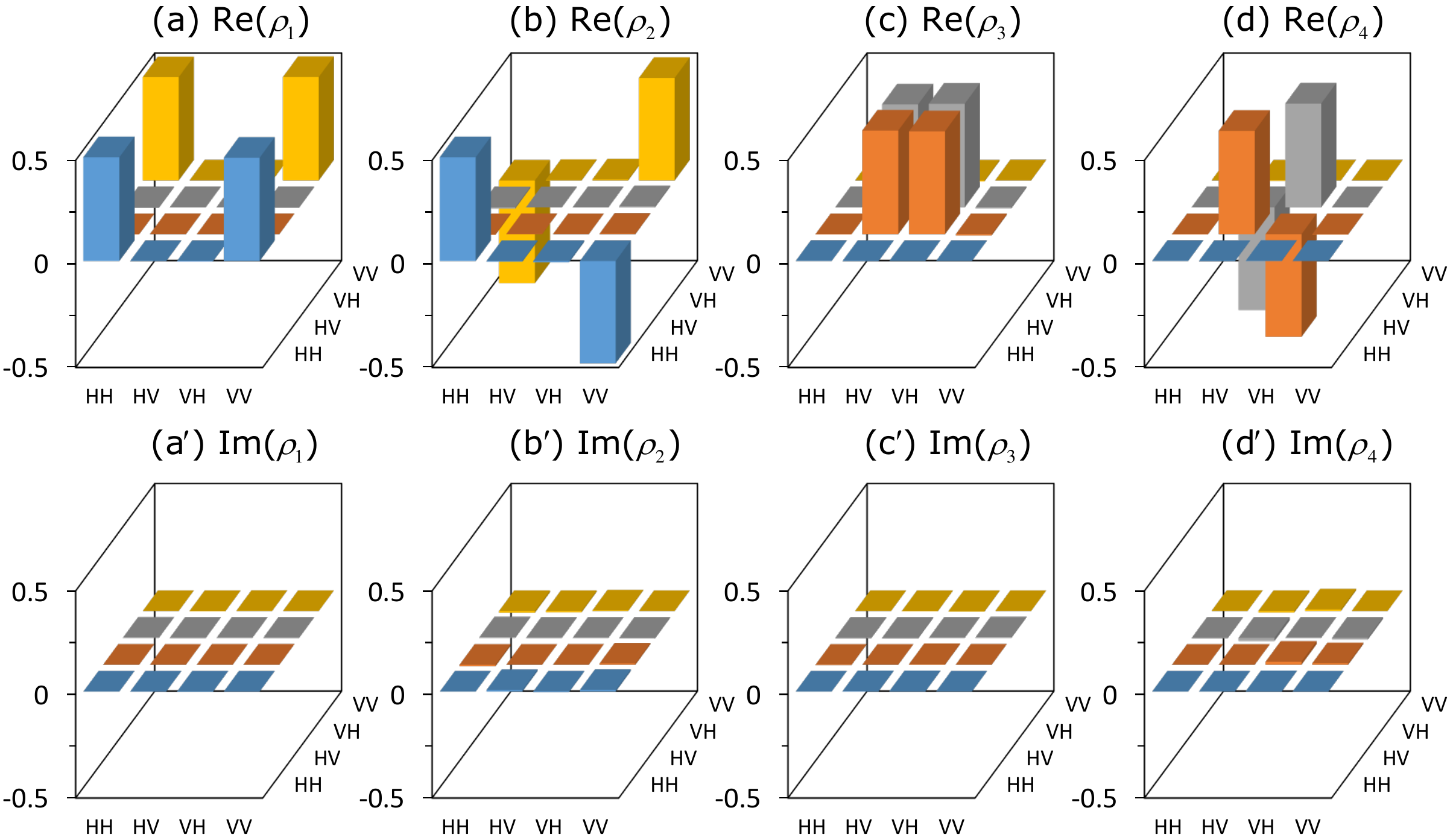}\\
			\caption{Tomographic results for the four maximally entangled states. (a) ${\mathop{\rm Re}\nolimits} ({\rho _1})$, (b) ${\mathop{\rm Re}\nolimits} ({\rho _2})$, (c)  ${\mathop{\rm Re}\nolimits} ({\rho _3})$ and (d) ${\mathop{\rm Re}\nolimits} ({\rho _4})$ in the upper plots represent real parts of $\rho _1$, $\rho _2$, $\rho _3$ and $\rho _4$, respectively. (a$'$) ${\mathop{\rm Im}\nolimits} ({\rho _1})$ , (b$'$) ${\mathop{\rm Im}\nolimits} ({\rho _2})$, (c$'$)  ${\mathop{\rm Im}\nolimits} ({\rho _3})$ and (d$'$) ${\mathop{\rm Im}\nolimits} ({\rho _4})$ in the lower plots represent imaginary parts of these states. }\label{Fig2}
		\end{figure}

		In order to visualize the tomographic results for the initial BDSs, we provide graphical representation of the reconstructed density matrix of the the four maximally entangled states: ${\rho _1}\;(p = 1,q = 1)$, ${\rho _2}\;(p = 0,q = 1)$, ${\rho _3}\;(p = 1,q = 0)$, and ${\rho _4}\;(p = 0,q = 0)$. Fig. \ref{Fig2} shows the real and imaginary parts of the four maximally entangled states, respectively. From the figure, one can see that they are four Bell states with high fidelity. Tomographic results for all the initial BDSs we have prepared are shown in the Supplemental Material.

		\begin{figure}
			\centering
			\includegraphics[width=8cm]{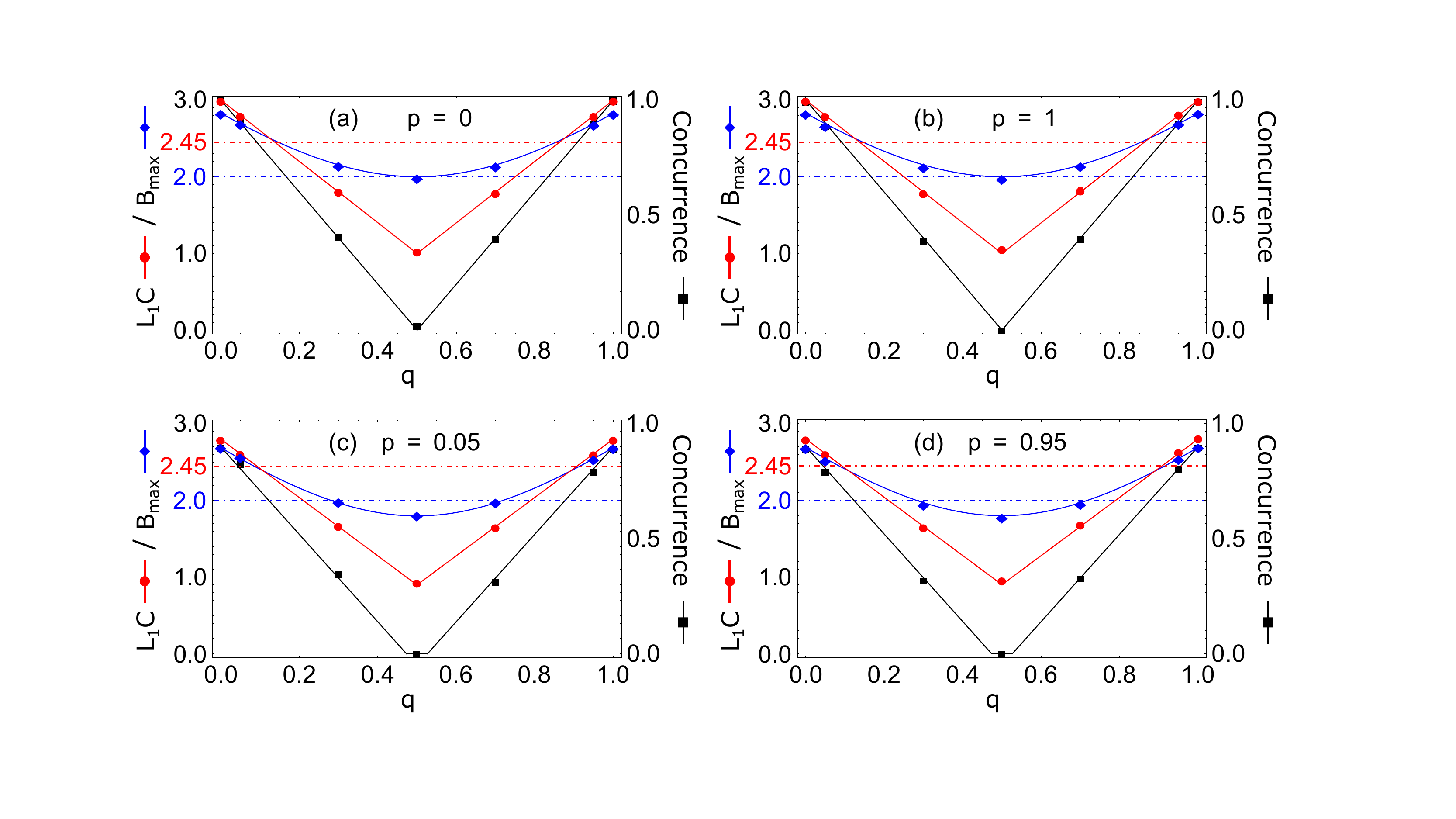}\\
			\caption{Experimental results and the corresponding theoretical predictions. The horizontal axis represents the parameter $q$ of the initial BDS. The vertical axis on the right represents concurrence, the vertical axis on the left represents the $l_1$ norm of coherence ($L_1C$) and the maximum violation of CHSH inequality. The black squares, blue rhombus and the red circles denote the experimental results of $C({\rho _{AB}})$, ${B_{\max }}({\rho _{AB}})$ and ${N_{{l_1}}}({\rho _{AB}})$. The solid lines in different colors are the corresponding theoretical predictions, respectively.}\label{Fig3}
		\end{figure}
		
		\begin{figure}
			\centering
			\includegraphics[width=8cm]{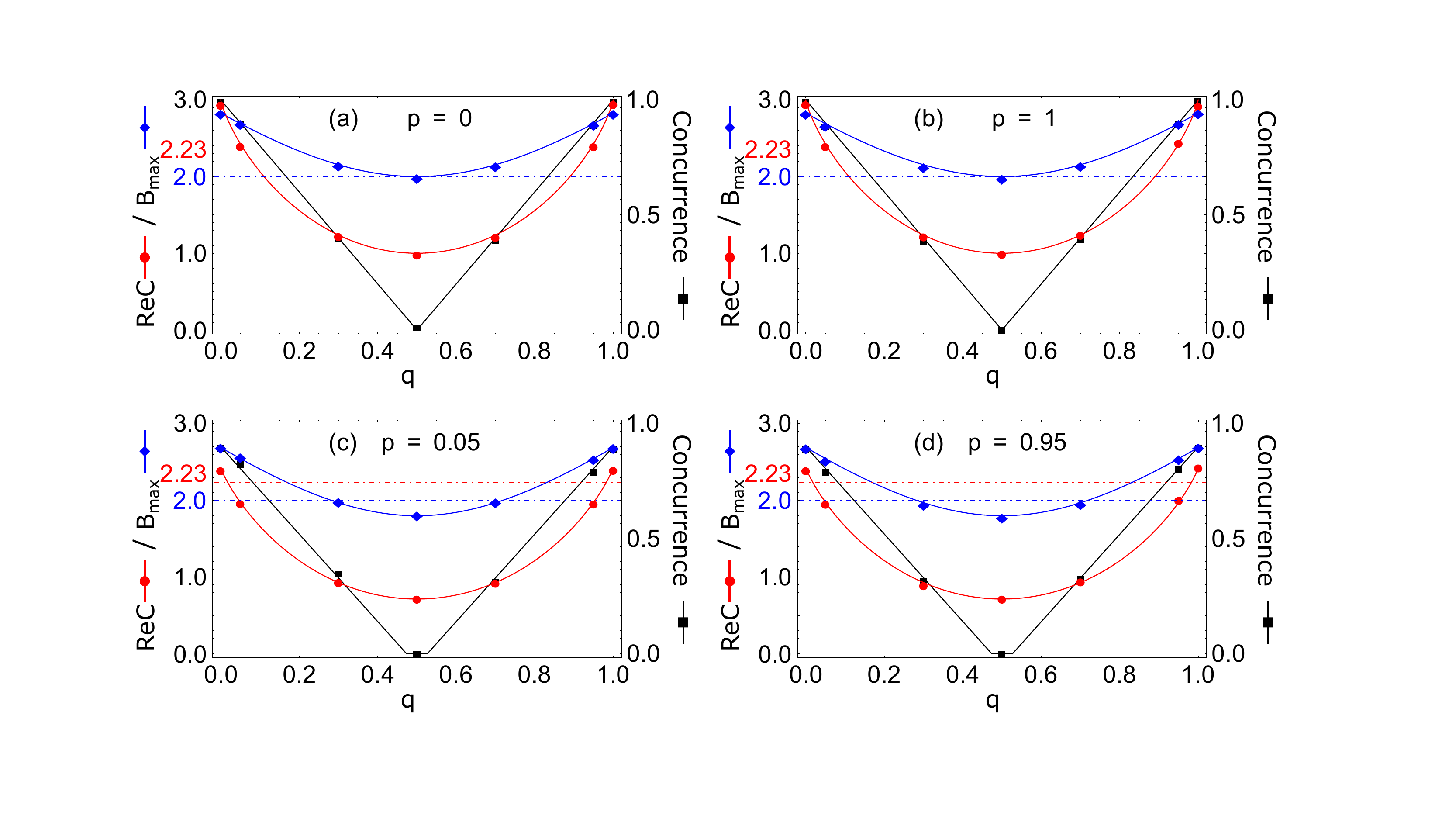}\\
			\caption{Experimental results and the corresponding theoretical predictions. Here, the vertical axis on the left represents the relative entropy of coherence (ReC). And the red circles denote the experimental results of ${N_{re}}({\rho _{AB}})$.}\label{Fig4}
		\end{figure}
		
		\begin{figure}
			\centering
			\includegraphics[width=8cm]{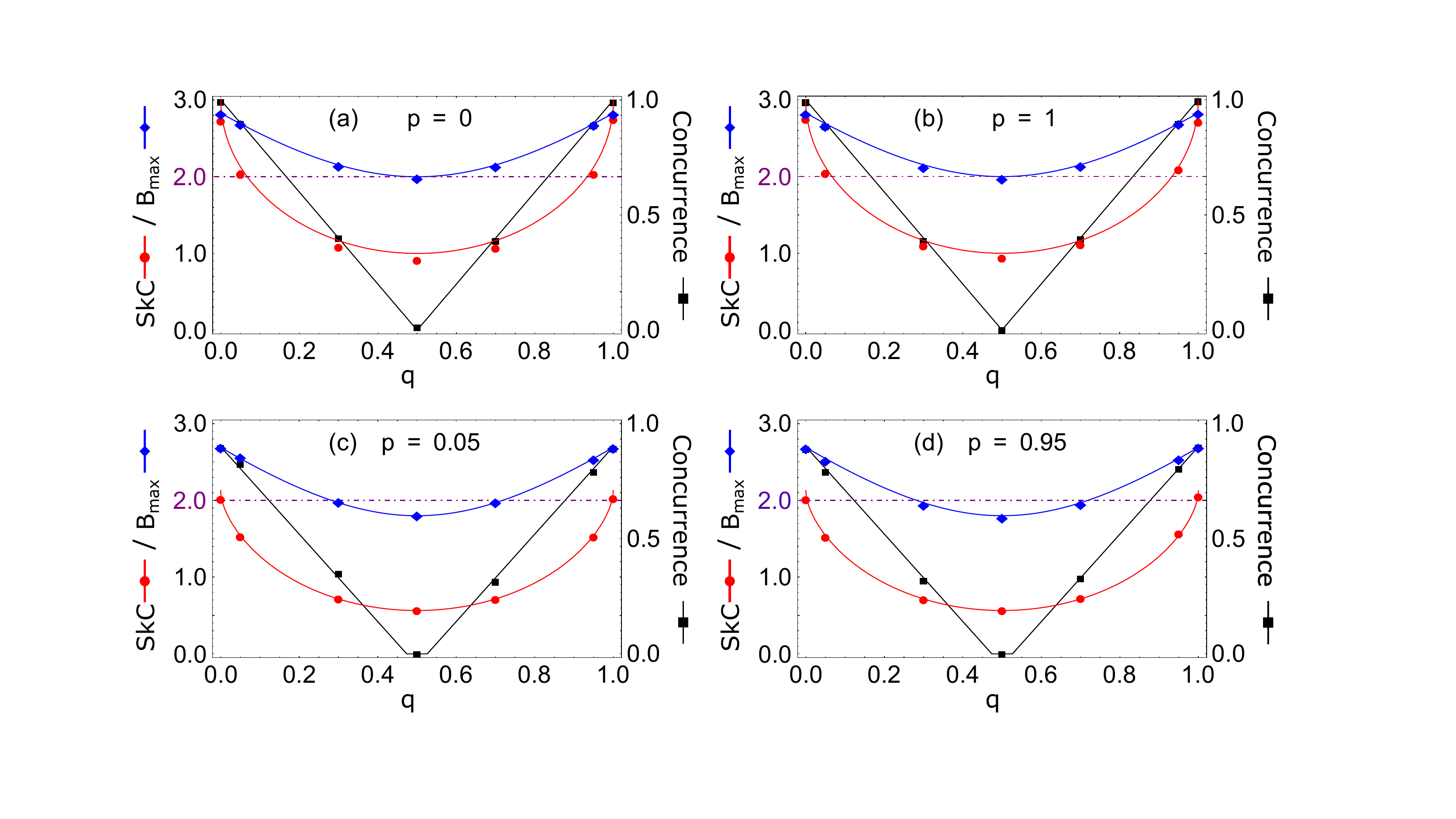}\\
			\caption{Experimental results and the corresponding theoretical predictions. Here, the vertical axis on the left represents the skew information of coherence (SkC), and the red circles denote the experimental results of ${N_{sk}}({\rho _{AB}})$.}\label{Fig5}
		\end{figure}
		
		The experimental results and the corresponding theoretical predictions have been shown in Figs. \ref{Fig3}-\ref{Fig5}. The horizontal axis represents the parameter $q$ of the initial BDS ${\rho _{AB}}$ in Eq. (7). The vertical axis on the right represents concurrence, the vertical axis on the left represents coherence and the maximum violation of CHSH inequality. The black squares and blue rhombus denote the experimental results of $C({\rho _{AB}})$ and ${B_{\max }}({\rho _{AB}})$, which are calculated from the quantum states ${\rho _{AB}}$ reconstructed by tomography. The red circles in Figs. \ref{Fig3}-\ref{Fig5} represent the experimental results of ${N_{{l_1}}}({\rho _{AB}})$, ${N_{re}}({\rho _{AB}})$ and ${N_{sk}}({\rho _{AB}})$  for the conditional state on Bob's side in Eq. (3), respectively. It should be noted that every experimental point in the figures has an error bar, but most of them are too small to display. The solid lines in different colors are the corresponding theoretical predictions. The red horizontal dot dash lines in Figs. \ref{Fig3}-\ref{Fig5} represent state-independent upper bound $C_{{l_1}}^m$, $C_{re}^m$ and $C_{sk}^m$. The experimental points above this horizontal line implies that the corresponding quantum states achieve NAQC. The blue horizontal dot dash lines represent the upper bound of Bell nonlocality. From these figures, the following conclusions can be obtained:
	
		(i) We capture NAQC for the initial BDSs based on the $l_1$ norm, relative entropy and skew information of coherence, and our experimental results coincide well with the theoretical predictions. NAQC is not only related to the form of quantum states, but also to the choice of coherence measures. For example, if we choose the $l_1$ norm as coherence measure, ${\rho _{AB}}\;(p = 0.05,q = 0.05)$ can achieve NAQC (see Fig. \ref{Fig3}c), but if we choose the other two coherence measure, this state cannot achieve NAQC (see Fig. \ref{Fig4}c and Fig. \ref{Fig5}c).
	
		(ii) It is obvious that quantum states with NAQC must violate the CHSH inequality, but not vice versa, which demonstrate that NAQC is a stronger nonclassical correlation than Bell nonlocality.
		
		(iii) It is worth mentioning that quantum states with NAQC would have higher entanglement, and the criterion of NAQC can also be considered as an indicator of entanglement. Therefore, NAQC is expected to be a kind of useful physical resource for quantum information processing.

	\section{CONCLUSIONS}
		In this paper, we have experimentally observed the NAQC for Bell-diagonal states with high fidelity in a completely optical setup. We perform local measurements on qubit $A$ in the three mutually unbiased bases and reconstruct the density matrices of the measured states by quantum state tomography process. And we also prepare the quantum states that hold inequality (3) based on the $l_1$ norm, relative entropy and skew information of coherence, respectively, which means that we prove the existence of NAQC by experimental methods. It can be seen that our experimental results coincide well with the theoretical predictions. It reveals that quantum states with NAQC would have higher entanglement. Thus, the criterion of NAQC can  be regarded as a good candidate of achieving entanglement witness. From the perspective of hierarchical relations, the NAQC can be considered as a stronger nonclassical correlation than Bell nonlocality, and thus can be expected to be a a useful physical resource in the regime of quantum information processing.
		
	\section*{ACKNOWLEDGMENTS}
		This work was supported by the National Science Foundation of China (Grant Nos. 11575001, 11405171, 61601002 and 11605028), the Natural Science Research Project of Education Department of Anhui Province of China under Grant No. KJ2018A0343, the Key Program of Excellent Youth Talent Project of the Education Department of Anhui Province of China under Grant Nos. gxyqZD2018065 and gxyqZD2019042, the Key Program of Excellent Youth Talent Project of Fuyang Normal University under Grant No. rcxm201804, the Open Foundation for CAS Key Laboratory of Quantum Information under Grant Nos. KQI201801 and KQI201804, the Research Center for Quantum Information Technology of Fuyang Normal University under Grant No. kytd201706.
		
		Zhi-Yong Ding and Huan Yang contributed equally to this work.
	
	\bibliographystyle{plain}

\end{document}